\documentclass[aps,pre,showpacs,twocolumn,floatfix]{revtex4}
\usepackage{bm}
\usepackage{graphicx}
\bibstyle{approve.bib}

\begin{document}
\title{Reparametrizing the loop entropy weights: effect on DNA melting curves}
\author{Ralf Blossey}
\affiliation{Interdisciplinary Research Institute c/o IEMN, Cit\'e
Scientifique BP 69, F-59652 Villeneuve d'Ascq, France}
\author{Enrico Carlon}
\affiliation{Interdisciplinary Research Institute c/o IEMN, Cit\'e
Scientifique BP 69, F-59652 Villeneuve d'Ascq, France}
\date{\today}

\begin{abstract}
Recent advances in the understanding of the melting behavior of
double-stranded DNA with statistical mechanics methods lead to improved
estimates of the weight factors for the dissociation events of the chains,
in particular for interior loop melting. So far, in the modeling of DNA
melting, the entropy of denaturated loops has been estimated from the
number of configurations of a closed self-avoiding walk. It is well
understood now that a loop embedded in a chain is characterized by
a loop closure exponent $c$ which is higher than that of an isolated
loop. Here we report an analysis of DNA melting curves for sequences of
a broad range of lengths (from $10$ to $10^6$ base pairs) calculated
with a program based on the algorithms underlying {\sc MELTSIM}. Using
the embedded loop exponent we find that the cooperativity parameter is
one order of magnitude bigger than current estimates. We argue that
in the melting region the double helix persistence length is greatly
reduced compared to its room temperature value, so that the use of the
embedded loop closure exponent for real DNA sequences is justified.
\end{abstract}

\pacs{87.14.Gg, 87.15.He, 05.70.Fh, 64.10+h}

\maketitle

\section{Introduction}

A standard method to unbind the two strands of a double-stranded
DNA (dsDNA) in solution is that of increasing the temperature of the
system. This process, known as DNA thermal denaturation or DNA melting,
has been studied since the sixties as it provides important information
about the interaction between base pairs, the stability of the double
helix, the effect of the solvent and of the salt concentration (for a
review see Ref. \cite{wart85}).

Several techniques are currently available for the investigation of
DNA thermal denaturation such as UV absorption, differential scanning
calorimetry, circular dichroism, NMR, fluorescence emission and
temperature gradient gel electrophoresis \cite{bloo00}. Perhaps the
most established of these is the UV absorption method which consists in
irradiating the sample with UV light at $270$ nm, a wavelength which is
preferentially absorbed by single stranded DNA (ssDNA). The total fraction
of absorbed light, $\widetilde A$, is therefore simply proportional to
the fraction of dissociated base pairs of the sequence and provides a
direct measurement of the order parameter of the problem. In experimental
studies, rather than analyzing directly $\widetilde A$, it is customary
to consider its temperature derivative: the plot of $d\widetilde{A}/dT$
vs. $T$ is known as the differential melting curve \cite{wart85}, but we
will simply refer to it, in the rest of the paper, as the melting curve.

\begin{figure}[t]
\includegraphics[height=4.3cm]{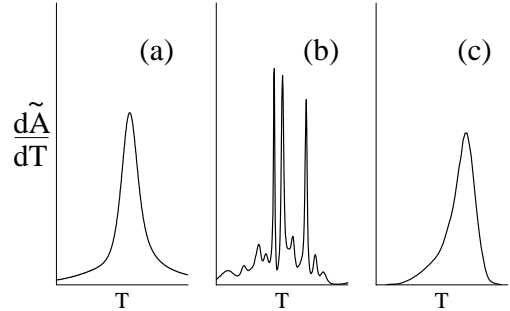}
\caption{
Schematic view of possible differential melting curves observed for (a)
short ($\approx 10^2$ bps), (b) intermediate ($\approx 10^3 - 10^4$ bps)
and (c) long ($> 10^6$ bps) DNA sequences.
}
\label{FIG01}
\end{figure}

Melting curves can show a single or several peaks (typical examples, as
observed in experiments, are shown in Fig. \ref{FIG01}) whose positions
and heights depend on the sequence length and composition as well as on
external parameters as the salt concentration. For very short sequences,
i.e. of about $10^2-10^3$ base pairs (bps), the melting curve shows
a single peak indicating a sudden unbinding of the two strands (see
Fig. \ref{FIG01}(a)). The peak is rounded by finite size effects and thus
can become quite broad for very short chains. Somewhat longer sequences
($\approx 10^3-10^4$ bps) are instead characterized by several peaks of
typical width of about $0.5^\circ C$ or less (Fig. \ref{FIG01}(b)). These
peaks are the signatures of sharp transitions of cooperatively melting
regions, as for instance inner loop openings or the unbinding of
double stranded regions at the edges of the chain. Finally, in very long
chains ($\approx 10^6$ bps) there is again only a single broad peak
covering about $15-20^\circ C$, which is actually the superposition of
many distinct peaks associated with the denaturation of single domains
[Fig. \ref{FIG01}(c)]. These single peaks cannot be resolved anymore
for such long sequences and the melting curve becomes again rather
featureless.

The computational prediction of DNA melting curves is a basic
bioinformatics task which is needed for a large variety of applications
such as primer design, DNA control during Polymerase Chain Reaction
or mutation analysis \cite{lyon01}. Also the denaturation behavior of
DNA sequences of genomic size is of interest for studies of sequence
complexity and evolution and for gene identification and mutation
\cite{yera00,yera00b,bizz98,gare03}. Consequently, specialized tools,
such as Meltsim \cite{blak99}, were developed for this purpose.

The first attempts to model the DNA denaturation dates back from the
sixties (for a recent review see Ref. \cite{theo03}).  The simplest
model employed for this purpose was the one-dimensional Ising model where the two
spin states $s_i = 0,1$ represent the open ($s_i = 0$) or closed ($s_i
=1$) configuration for complementary base pairs \cite{zimm59,zimm60}.
This model obviously does not provide a real thermodynamical phase
transition but only a smooth crossover between the closed (dsDNA) and
open (ssDNA) regimes. The inclusion of an entropic term, which takes
into account the number of possible configurations of the denaturated
loops, induces an effective long range interaction in the system and
thus a genuine phase transition may occur, as shown long ago by Poland
and Sheraga \cite{pola66}.

On rather short DNA sequences ($\sim 10^2$ bps) the loop entropy
contribution is not very important as loops are rare and rather
short and the DNA denaturates mainly trough unbinding from the edges.
A description based on the 1d Ising model with appropriate experimentally
determined energy parameters is therefore sufficient (see, e.g.,
Ref. \cite{owcz97}).  A calculation aimed at reproducing experimental
melting curves with several peaks of different heights and widths [as
that shown in Fig. \ref{FIG01}(b)] however needs, besides accurate energy
parameters, also a good estimate of the entropy of the denaturated loops.

A broadly known method to calculate DNA melting curves for a given input
sequence is the Poland algorithm \cite{pola74} in which the probability
for each base pair to be in an open or closed state can be calculated from
a set of recursive relations.  The entropy of the denaturated loops is
given by counting the number of configurations for a closed self-avoiding
walk \cite{fish66}. This approach overestimates the actual entropy as it
does not take into account that the number of configurations available
to the loop is limited by the presence of the rest of the chain. Recent
advances in the statistical mechanics of polymers provided new insight
on how to calculate entropies for loops which are embedded in chains
\cite{kafr00,carl02}.

The aim of this paper is to discuss whether DNA thermal denaturation
experiments are able to fix unequivocally the form of the parameters
involved in the entropy of denaturated loops. We analyze the effect of the
improved loop entropy estimate on the melting curves for DNA sequences of
various lengths (up to $5 \times 10^6$ base pairs) and compositions. We
show that the modification of the entropic parameters associated with the
loops may have quite a strong effect on the melting curves, especially
for sequences of intermediate lengths where several peaks are present
[as in the example of Fig. \ref{FIG01}(b)]. We also discuss how the
effects described here can be best measured experimentally.

This paper is organized as follows: In Sec. \ref{sec:entropy} we review
some recent results on the entropy of loops embedded into a chain. In
Sec. \ref{sec:meltsim} we present some melting curves for sequences
of various lengths and investigate the effect of a modification
of the parameters associated with the loop entropy on these curves.
In Sec. \ref{sec:stiffness} we discuss how the double helix rigidity could
influence the results for the entropy, while Sec. \ref{sec:conclusion}
concludes the paper.

\section{Entropy of a loop embedded in a chain}
\label{sec:entropy}

\begin{figure}[b]
\includegraphics[height=2.3cm]{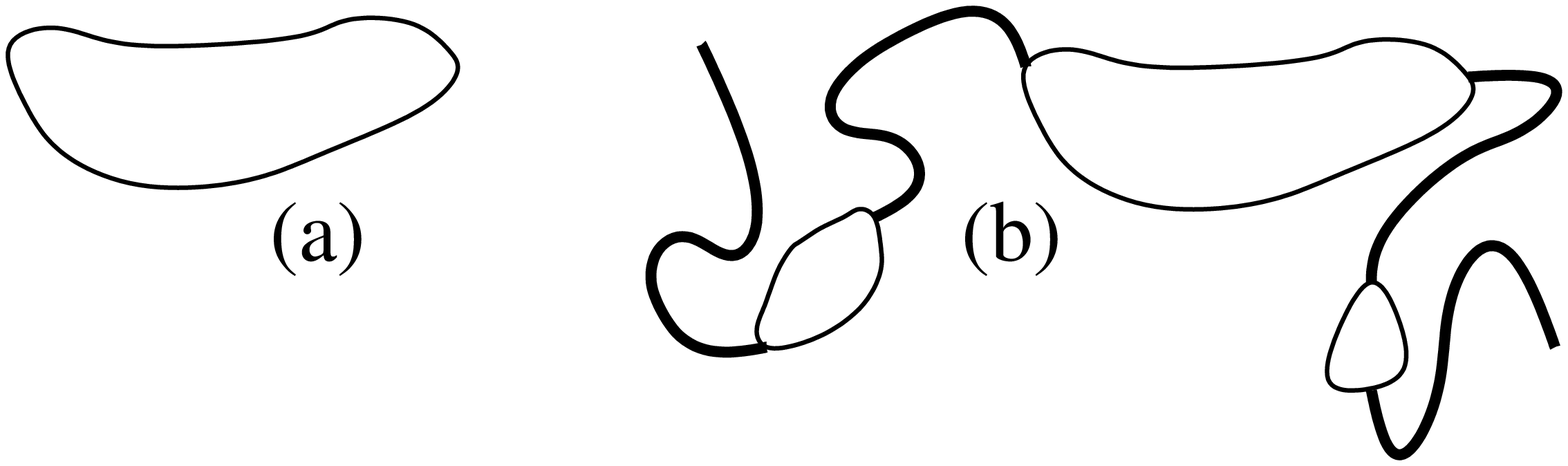}
\caption{
The total number of configuration for a self-avoiding loop is given by
Eq. (\ref{loop_entropy}) with $c \approx 1.75$ for an isolated loop (a)
and $c \approx 2.15$ for a loop embedded in a chain (b).
}
\label{FIG02}
\end{figure}

In the Poland algorithm the partition function of a loop of total length
$l$ is given by the number of configurations of a self-avoiding walk
returning for the first time to the origin \cite{fish66} after $l$ steps
(Fig. \ref{FIG02}(a)), which in the limit $l \to \infty$ assumes the
following form \cite{vand98}:
\begin{equation}
L_l \sim \sigma \frac{\mu^l}{l^c}
\label{loop_entropy}
\end{equation}
where $\mu$ is a nonuniversal geometric factor, while $c$, the so
called loop closure exponent, is a universal quantity. In Eq.
(\ref{loop_entropy}) we also included a prefactor $\sigma$, which
only makes sense when comparing the loop partition function with
that of a double stranded helix, and measure the absolute
probability of interrupting a double helix to open a loop. This
quantity is known as the {\it cooperativity parameter}.

A small value of $\sigma$ suppresses loop formation so that loops
proliferate only at temperatures very close to the melting point and are
typically large in order to minimize the effect of a small $\sigma$.
The transition becomes highly cooperative, in the sense that large
portions of the chain will tend to unbind simultaneously.  On the
contrary, for a $\sigma \approx 1$, there is no extra big penalty for
loop openings and many small loops may form already well below the
melting point. The transition is less cooperative, and peaks in the
melting curves appear more rounded-off.

For self-avoiding walks embedded in a three dimensional space the
exponent $c$ has been estimated numerically to be $c \approx 1.75$
\cite{vand98}. Other loop parameters have to be fixed by fitting
to the available experimental data. In particular, a lot
of effort has been devoted to the measurement of the cooperativity
parameter $\sigma$. Its most accurate determination was performed by
Blake and Delcourt \cite{blak98} who analyzed the melting of several
tandemly repeated inserts on a linearized plasmid DNA and found that
the value $\sigma = 1.26 \times 10^{-5}$ fits best the experimental
melting curves. This value is consistent with previous estimates
\cite{wart85,oliv77,amir81,blak87}.

Recent developments in the field of polymer physics allow us to calculate
the total number of configurations for a loop embedded in a chain, as
for those shown in Fig. \ref{FIG02}(b). Analytical estimates, relying
on the general theory of polymer networks \cite{dupl86}, indicate that
the form given in Eq.  (\ref{loop_entropy}) is still valid, however
with a higher value for the exponent $c$ \cite{kafr00}. The reason
for the higher value of $c$ is the lower number of configurations
available for the loop due to the presence of the rest of the chain,
compared to the number of configurations available for an isolated loop.
Monte Carlo simulations on suitable three dimensional lattice models
yield $c \approx 2.15$ \cite{carl02,baie02}, a value which is also
in agreement with analytical estimates, which place this exponent in
the range $2.10 \lesssim  c \lesssim 2.20$ \cite{kafr00}. It is also
well-known \cite{pola66,fish66} that increasing the value of $c$ has an
effect of sharpening the transition. A value $c > 2$ implies a first order
transition in the case that the energy difference between different base
pairs is neglected \cite{pola66,fish66}; first order melting was indeed
found numerically \cite{caus00}.

It is clear that accepting $c=2.15$ as the most appropriate estimate of
the loop exponent entering in the partition function (\ref{loop_entropy})
implies that the existing estimate of the cooperativity parameter $\sigma$
has to be revisited as its experimental determinations relied on the
choice $c=1.75$. The aim of this paper is to investigate this issue in
detail. To this purpose we analyze the melting curves for DNA sequences
of a wide range of lengths up to the full genome of the E. coli which
amounts to $5 \times 10^6$ bps.

\section{Calculation of melting curves}
\label{sec:meltsim}

The calculations of DNA melting curves are obtained from our own
C-version of the Meltsim package (for details concerning this program,
see Ref. \cite{blak99}), which is a program based on the Poland recursive
algorithm for the calculation of the probabilities $A(i)$ that the
$i$-th base pair is in a open state.  The program uses the Fixman-Friere
\cite{fixm77} method, which consists of approximating the loop partition
function of Eq.  (\ref{loop_entropy}) as a sum of $I$ exponentials:
\begin{equation}
L_l \approx \sigma \sum_{k=1}^I a_k e^{-b_k l}
\end{equation}
where $a_k$ and $b_k$ are fitting coefficients. As for a sequence of
total length $N$ the longest possible loop corresponds to $l = 2N$,
the fitting is done in a limited interval of lengths.  Therefore,
the shorter the sequence, the smaller is the number of coefficient $I$
necessary to obtain a very good fit of Eq.  (\ref{loop_entropy}) for
the relevant loop lengths for the problem. Typically a sum with $I
\approx 10$ coefficients is sufficient for our needs. The advantage
of the exponential approximation is that it reduces the computational
time for a sequence of $N$ base pairs from $O(N^2)$ to $O(N \times I)$
without any significant loss in accuracy \cite{fixm77}.

For various temperatures we calculated $A(i)$, the probability that the
$i$-th base pair is in an open state. From numerical differentiation
of the average: 
\begin{equation} 
A = \frac 1 N \sum_{i=1}^N A(i)
\end{equation} 
for a sequence of $N$ base pairs, we obtained the melting curves $dA/dT$
vs. $T$.

In this work we have chosen the ten stacking energy parameters given in
Ref. \cite{blak99}. Another parameter that needs to be specified in
the calculation is the concentration of the monovalent salt in solution,
which in our calculation varies between $0.05$ M and $0.1$ M.

\begin{table}[t]
\caption{Domains from pN/MCSx from Ref. \cite{blak98}
considered here as insertions into purely CG-domains.}
\begin{ruledtabular}
\begin{tabular}{ccc}
Seq. (x) & Length (N) & Composition\\ \hline
 1 & 155 & [ACTCGGACGA]$_{15}$ACTCG\\
 2 & 305 & [ACTCGGACGA]$_{30}$ACTCG\\
 5 & 335 & [ACTCGGACGA]$_{33}$ACTCG\\
10 & 500 & [AAGTTGAACAAAT]$_{38}$AAGTTG\\ 11 & 747 &
[AAGTTGAACAAAT]$_{57}$AAGTTG\\ 12 & 214 &
[AAGTTGAACAAAT]$_{16}$AAGTTG\\ 13 & 245 &
[AAGTTGAACAAAAT]$_{17}$AAGTTGA\\ 14 & 203 &
[AAGTTGAACAAAAT]$_{14}$AAGTTGA\\
 3 & 135 & [AAGTTGAACAT]$_{13}$AAGTTG\\
 6 & 330 & [AAGTTGAACAAT]$_{27}$AAGTTG\\
15 & 292 & [AGTGACAT]$_{36}$AGTG
\end{tabular}
\end{ruledtabular}
\label{tableI}
\end{table}

\begin{figure}[t]
\includegraphics[scale=0.45]{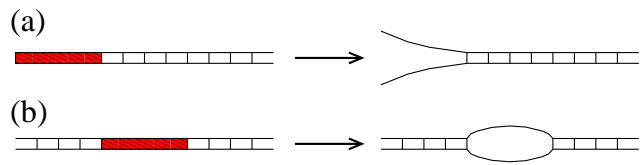}
\caption{
Schematic view of the melting of the AT-rich insert (gray) in the
case of (a) edge insertion and (b) inner insertion as performed in
Ref. \cite{blak98}. The difference in the melting temperatures for the
two cases $\Delta T_{\rm M} = T_{\rm M}^{\rm loop} - T_{\rm M}^{\rm end}$
allows to estimate of the cooperativity parameter.
}
\label{FIG03}
\end{figure}

\subsection{Melting of short sequences}

We first analyzed a series of existing experimental data \cite{blak98}
for short sequences consisting of oligomeric repeat patterns shown
in Table \ref{tableI}. These sequences were inserted in a linearized
recombinant pN/MCS$x$ plasmid ($x$ abbreviates a specific sequence). The
insertions are AT-rich and melt at lower temperatures than the rest
of the plasmid chain, which plays the role of an energetic barrier to
prevent further melting. The insertions were placed at one end, and in
the middle of the linearized plasmid, thus their melting occurs through
end or loop openings, respectively (see Fig.  \ref{FIG03}).

The experimental results, tabulated in Ref. \cite{blak98}, are
re-analyzed here using the exponent $c=2.15$, appropriate for a
loop inserted in a chain.  To induce both loop and end openings we
embedded the sequences of Table \ref{tableI} in the middle and at
one end of a pure CGCGCG chain. Due to the higher stability of the
CG-domains, their melting will occur at much higher temperatures
than those of the inserted sequences.

\begin{figure}[t]
\includegraphics[width=7cm]{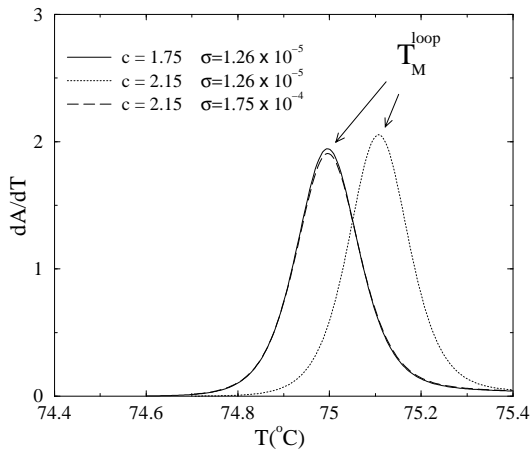}
\caption{
Calculated melting curves for the Seq. 11 inserted in the middle of a
long CGCG\ldots matrix, for $0.075$ M of monovalent salt concentration
and three choices of $c$ and $\sigma$.
}
\label{FIG04}
\end{figure}

Figure \ref{FIG04} shows a plot of the differential melting curves
for the Seq. 11 of Table \ref{tableI}, embedded in the middle of a
CG matrix, as calculated by our program. We first fixed $\sigma =
1.26 \times 10^{-5}$, the value reported in \cite{blak98}. The change
of $c$ from $1.75$ to $2.15$ causes a shift of the position of the
melting curves peak of about $0.2 ^\circ$ C and a slight increase of
height (the peak position defines the loop melting temperature $T_{\rm
M}^{\rm loop}$). By choosing $c= 2.15$, $\sigma=1.75 \times 10^{-4}$ one
recovers a melting curve which almost perfectly overlaps the original one
obtained with $c= 1.75$, $\sigma=1.25 \times 10^{-5}$ (solid and dashed
lines of Fig. \ref{FIG04}). This curve fits well the experimental data
(see Ref. \cite{blak98}), thus we conclude that for the choice $c=2.15$
and $\sigma=1.75 \times 10^{-4}$ yields a melting curve consistent with
the experimental values.

\begin{figure}[t]
\includegraphics[width=7cm]{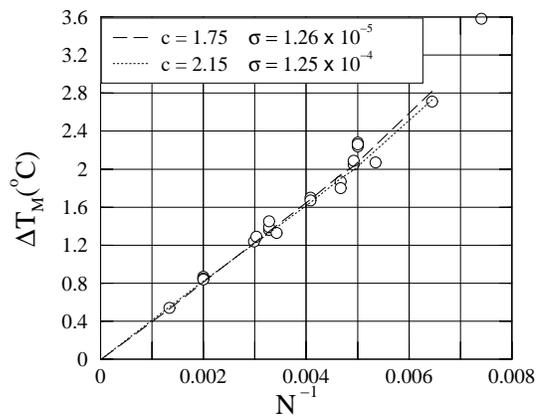}
\caption{
Plot of $\Delta T_M$ as function of the inverse domain size. Empty
circles are experimental data taken from \cite{blak98} (Table I)
dashed and dotted lines refer to two choices of $c$ and $\sigma$.
}
\label{exp_deltaT}
\end{figure}

In order to provide an estimate of $\sigma$ from several independent
measurements we reanalyzed the procedure followed in Ref. \cite{blak98}.
Figure \ref{exp_deltaT} shows experimental data (empty circles) for
the temperature difference of the sequences of Table \ref{tableI},
$\Delta T_M \equiv T_M^{\rm loop} - T_M^{\rm end}$, plotted as functions
of the inverse domain length $1/N$ (data taken from Table I of
Ref. \cite{blak98}). Here $T_M^{\rm loop}$ and $T_M^{\rm end}$ denote
the location of the maxima of $d A/ dT$ for sequences inserted in the
interior and at the end of the plasmid chain, The dashed line shows the
calculated $\Delta T_M$ in the case of $c=1.75$ $\sigma = 1.26 \times
10^{-5}$, where the latter value was determined using a regression
analysis to fit the experimental data. We have repeated this procedure
here fixing $c= 2.15$ for which we find an equally good fit of the data
with the choice $\sigma=1.25 \times 10^{-4}$. Given the precision of the
experimental data which we could not assess or analyze further, we find
that our calculated curve matches the data very well.  Deviations between
both theoretical curves clearly appear for shorter chains (loops) where
the application of the asymptotic form of the loop partition function
of Eq. (\ref{loop_entropy}) may not be appropriate.

\subsection{Melting of sequences of intermediate length}

We next consider the melting of two sequences of intermediate length. We
used two protein-coding cancer-related genes, eIF-4G ($2900$ bps) and
LAMC1 ($7900$ bps), selected from a series of other sequences analyzed
for their predominant loop melting effects.

\begin{figure*}[t]
\includegraphics[height=6.0cm]{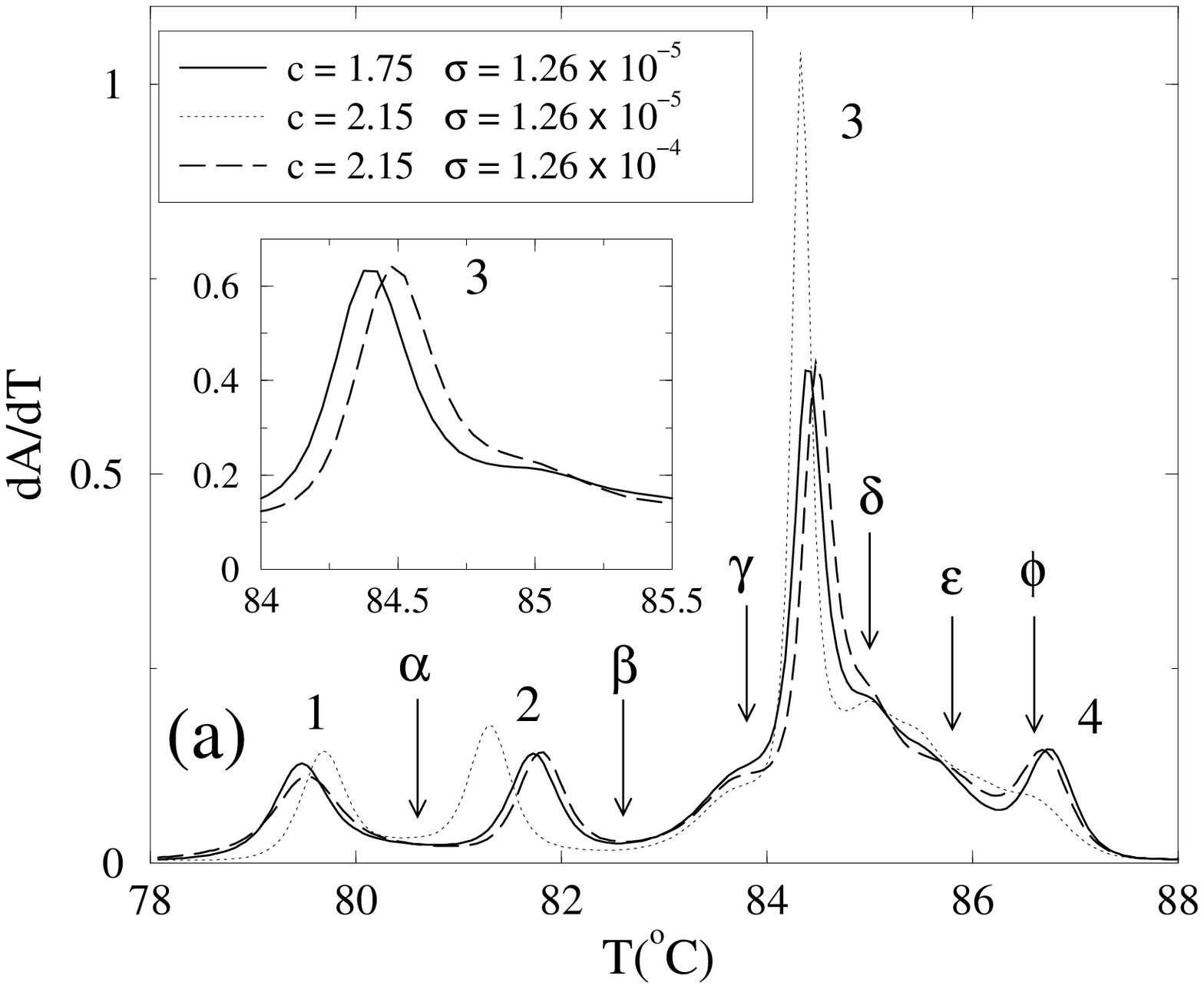}
\ \ \ \ \ \
\includegraphics[height=6.0cm]{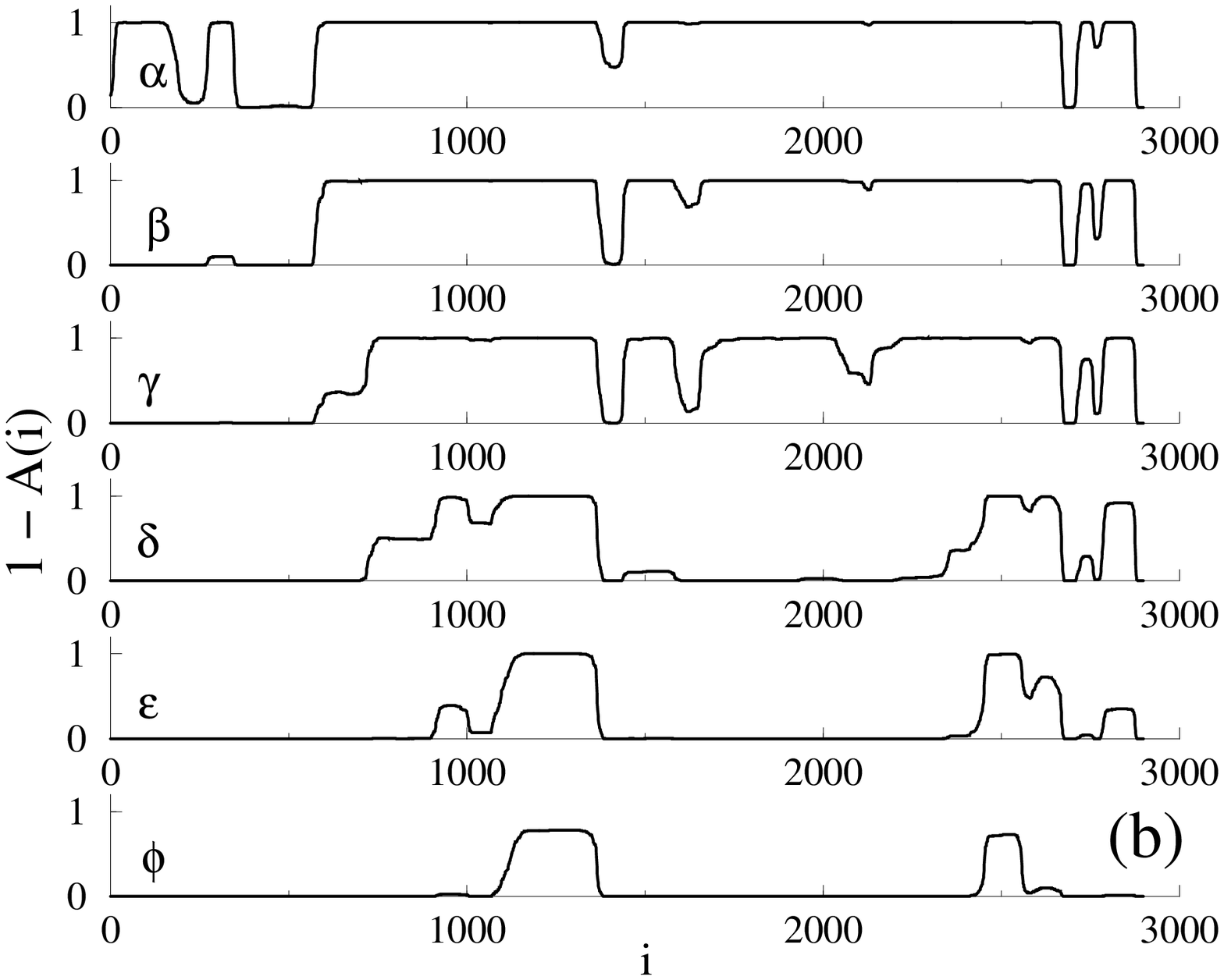}
\caption{(a) Melting curves for the sequence eIF-4G for three choices
of $c$ and $\sigma$ and $0.05$ M of monovalent salt. (b) Denaturation
maps for $c=2.15$, $\sigma=1.26 \times 10^{-5}$ calculated at the six
different temperatures (labeled by $\alpha$, $\beta$ \ldots ) indicated
by vertical arrows in (a).}
\label{eifg4}
\end{figure*}

Figure \ref{eifg4}(a) shows the melting curves for three different values
of $c$ and $\sigma$ for a fragment of DNA of eIF-4G as calculated from our
program. Four major distinct peaks, labeled from $1$ to $4$ are visible.
The {\it denaturation maps}, i.e.  plots of $1-A(i)$ as function of $i$,
the base position along the chain, which are shown in Fig. \ref{eifg4}(b),
provide further insight on the type of transitions associated with each
peak. We recall that $1-A(i)$ is the average probability that the $i$-th
base pair is in a closed state. The six plots of Fig.  \ref{eifg4}(b),
labeled as $\alpha$, $\beta$ \ldots $\phi$, correspond to the six
temperatures marked by the vertical arrows in Fig. \ref{eifg4}(a).

\begin{figure*}[t]
\includegraphics[height=6.0cm]{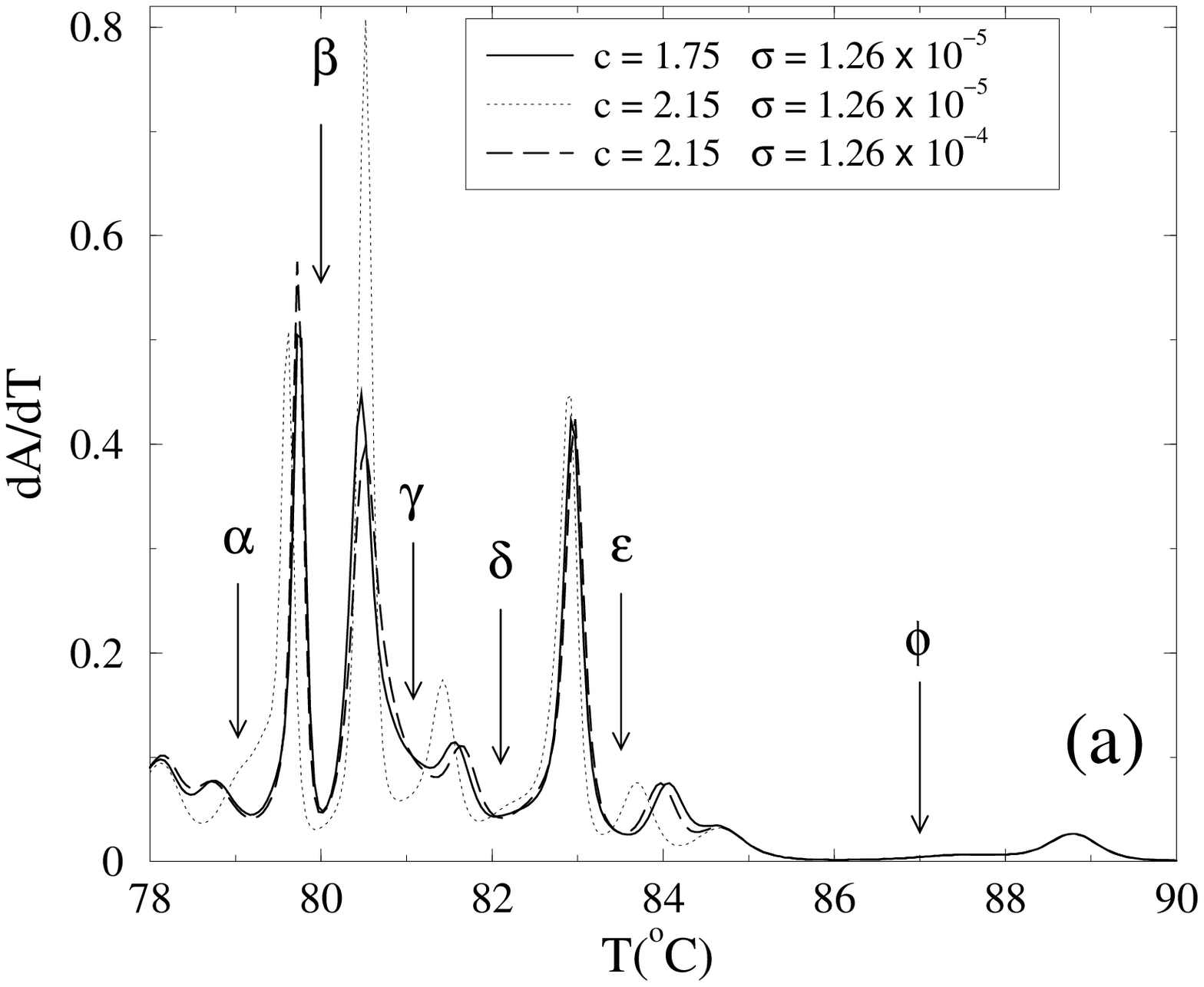}
\ \ \ \ \ \
\includegraphics[height=6.0cm]{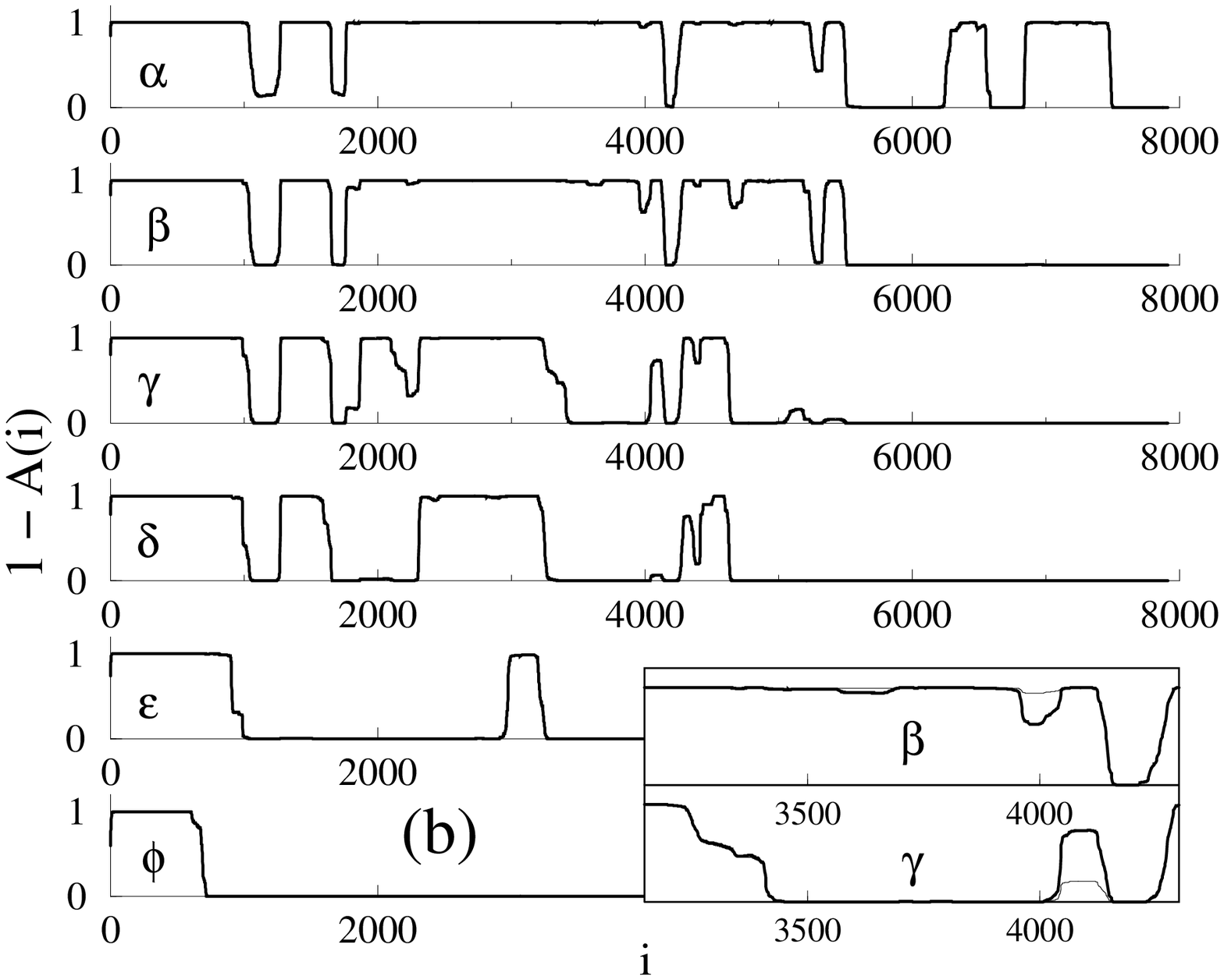}
\caption{(a) Plot of the melting curves for the DNA sequence of
LAMC1 for three choices of the parameters $c$ and $\sigma$. (b)
Denaturation maps for $c=2.15$, $\sigma = 1.26 \times 10^{-4}$ at
the six different temperatures marked in (a) by the vertical
arrows. Inset: Blow up of the denaturation maps $\beta$ and
$\gamma$ for $c=2.15$, $\sigma = 1.26 \times 10^{-4}$ (thick solid
lines) and $c=2.15$, $\sigma = 1.26 \times 10^{-5}$ (thin solid
lines).} 
\label{LAMC1}
\end{figure*}

Comparing the configurations at the temperatures just below and
above the peak $3$ ($\gamma$ and $\delta$) we notice that this
peak corresponds to the opening of a big loop extending roughly
from base pair $1300$ to $2300$.  The melting curves were
calculated for three different sets of parameters, starting from
the standard choice $c=1.75$, the value associated with isolated
loops, together with $\sigma = 1.26 \times 10^{-5}$, which is the
most recent estimate for the cooperativity parameter
\cite{blak98}. While keeping $\sigma$ fixed we first consider the
exponent for a loop embedded in the chain $c=2.15$. For such a
combination of parameters the melting peaks $1$ and $2$ are
shifted from their positions, while peak $4$ vanishes in the
signal background. The main influence of a change in $c$ is on
peak $3$, which for $c=2.15$ becomes roughly twice as high
($dA/dT|_{\rm max} \approx 1.1$ compared to $dA/dT|_{\rm max}
\approx 0.6$ in the case $c=1.75$). The reason for this is that,
as shown from the study of the denaturation maps, the peak $3$
corresponds to the opening of a long loop of about $1000$ bps,
thus any modification of the parameters entering in the loop
entropy will have a particularly strong effect on it. Increasing
$c$, while keeping $\sigma$ fixed makes the transition sharper, as
expected \cite{pola66,fish66}.

The third melting curve we plotted is obtained again with $c=2.15$, but
with a cooperativity parameter which is one order of magnitude larger:
$\sigma = 1.26 \times 10^{-4}$. This curve overlaps quite well with the
original one. The region where the first and third curves differ the
most is around peaks $3$ and $4$, where the chain contains a $1000$
bps long loop, and is shown in the inset of Fig. \ref{eifg4}(a).  Even in
this region the differences are not very big: e.g., the shift of
the maximum for peak $3$ is $\Delta T_{\rm max} \approx 0.1^\circ$,
while both peaks keep the same heights.

We repeated similar calculations for other sequences in the same range of
lengths. Figure \ref{LAMC1}(a) shows the melting curves for a fragment
of LAMC1 about $7900$ bps long, for the same choice of parameters as in
Fig. \ref{eifg4}. Compared to the previous example there is a larger
number of peaks, as the sequence is more than twice as long as the
previous one and more subtransitions take place.  The most relevant
difference between the melting curves calculated for different values
of $c$ and $\sigma$ is within the region around $81^\circ C$, where
a melting peak doubles its height going from the original choice of
$c=1.75$ to $c=2.15$, while keeping the cooperativity parameter fixed at
$\sigma = 1.26 \times 10^{-5}$. As in the eIF-4G sequence, when $\sigma$
is rescaled by a factor $10$ we obtain a melting curve running extremely
close to the original one in the whole range of temperatures.

It is interesting to take a closer look at the average configurations
around the peak at $T=80.5^\circ C$. The inset of Fig. \ref{LAMC1}(b)
shows a blow up of the denaturation maps $\beta$ and $\gamma$
in a region around base pairs $i=3500-4000$, where a loop opening
occurs. While keeping $c=2.15$ we plot in the inset the maps both for
$\sigma = 1.26 \times 10^{-5}$ (thin lines) and $\sigma = 1.26 \times
10^{-4}$ (thick lines).  It is remarkable how little thick and thin
lines differ, compared to the big effect of a change of $\sigma$ in the
differential melting curves of Fig.  \ref{LAMC1}(a). The "robustness"
of the denaturation maps with respect to a change of the thermodynamic
parameters has also been observed recently by Yeramian \cite{yera00}. The
curves in the inset clearly demonstrate the increase of cooperativity
when $\sigma$ is decreased: Either loops are suppressed (case $\beta$) or
two neighboring loops tend to merge (case $\gamma$) in order to minimize
the effect of a small value of $\sigma$. Although the peak at $83^\circ$
C corresponds to the formation of a loop of about $2000$ bps, as can be
seen from the denaturation maps $\delta$ and $\varepsilon$, a change
in the loop parameters $c$ and $\sigma$ does not seem to modify much
this peak, as the transition considered is not an opening of a double
helical segment, but rather a merging of two loops with a corresponding
enlargement. We conclude that loops merging transitions are less affected
by a change in $c$ and $\sigma$ as compared to a genuine loop opening.

We notice also that changing the loop parameters do not affect at
all the small peak at $T \approx 89^\circ C$, because this corresponds
to a transition not involving inner loops, but rather the melting of
the region $0 < i < 900$ through opening from the edges (see $\phi$
in Fig. \ref{LAMC1}(b)).

\begin{figure}[b]
\includegraphics[height=6.0cm]{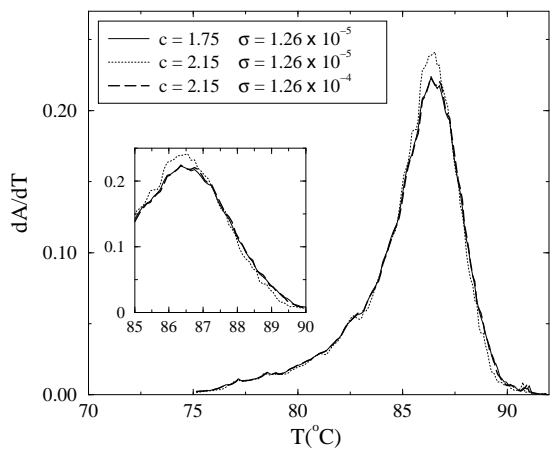}
\caption{Melting curves for the whole DNA for E. coli ($\approx
4,500,000$ bps).} \label{ecoli_fig}
\end{figure}

\subsection{Melting of long sequences (E-Coli)}

In order to further asses the validity of the previous analysis we
calculated melting curves for much longer DNA sequences ($\sim 10^6$
bps). Longer sequences have the advantage that they develop loops of
a broad range of length scales, thus one may test the influence in a
change of loop parameters simultaneously for very short and very long
loops. Unfortunately a limiting factor in this case is that, as pointed
out in the introduction, the melting curves are rather featureless
(as for Fig. \ref{FIG01}(c)).

Figure \ref{ecoli_fig} show the calculated melting curves for the full DNA
of E. coli which is of about $4.6 \times 10^6$ bps with a known sequence
(taken from \cite{ecoli}).  The curves are almost continuous, although
some irregular structure is still visible, indicating that the discrete
sharp peaks of the underlying sequence have not been completely averaged
out. The melting curves extend now over about $15^\circ C$ and have a
typical asymmetric shape, with a gentle growth in the low temperature
region, and a steeper descent above $T_{\rm max}$, the temperature for
which $dA/dT$ is maximal.

As before we calculate the melting curves for three different combinations
of the parameters $c$ and $\sigma$. Again we find that by changing
the loop closure exponent from $c=1.75$ to $c=2.15$ while rescaling of
$\sigma$ by one order of magnitude brings the curve back to the original
one.  The two curves obtained in this way are perfectly overlapping. A
change of $c$ only, while leaving $\sigma = 1.26 \times 10^5$ makes
the melting curve somewhat sharper as observed previously for shorter
sequences. However the increase of the peak height is limited to about
$10 \%$, a rather small effect compared to the doubling of the height
found for some peaks of Figs. \ref{eifg4} and \ref{LAMC1}.

It is interesting to notice that by changing the loop parameters $c$
and $\sigma$ the part of the melting curves for $T \lesssim T_{\rm
max}$ apparently is not modified much as all three curves plotted in
Fig. \ref{ecoli_fig} run with good accuracy on top of each other, in this
temperature interval. It is only at at $T \approx T_{\rm max}$ and above
that the curves start separating, as shown in the inset. We also point
out that the comparison of experimental and calculated melting curves
for E. coli of Ref.  \cite{blak99} show some differences in the high
temperature region $T \gtrsim T_{\rm max}$. These differences could be
due to some non-equilibrium effects, which are known to be more severe
for very long sequences \cite{wart85}.

\section{Effect of the double helix stiffness}
\label{sec:stiffness}

There has been some discussion recently about the influence of the
stiffness of the double helix on the exponent $c$
\cite{hank03,kafr03}. In the very ideal limit of a loop attached
to two infinitely rigid rods, the appropriate value of the loop
exponent is the same of that of an isolated loop $c=1.75$, as the
loop does not interact with the rest of the chain. The dsDNA has a
persistence length which is typically much larger than that of the
ssDNA  ($\xi_{\rm ds} \gg \xi_{\rm ss}$), therefore it is
legitimate to question the applicability of the polymer network
theory \cite{dupl86}, from which a higher loop exponent $c \approx
2.1$ was derived \cite{kafr00}, to real DNA sequences.

A different persistence length between dsDNA and ssDNA was incorporated
in lattice Monte Carlo simulations \cite{carl02}. In these calculations
the exponent $c$, estimated from an analysis of the distributions of the
loop lengths using realistic values for $\xi_{\rm ds}$ and $\xi_{\rm ss}$,
was found to be consistent with that of the case in which the difference
between persistence lengths is neglected. However these lattice models
did not incorporate a cooperativity parameter.

In Ref. \cite{hank03} it was estimated that sequences up to $5000$
bps are still too short to show any interaction effects between a loop
with the rest of the chain, so that the appropriate value of the loop
exponent should be that of an isolated loop $c \approx 1.75$.  There is
however still disagreement about this conclusion \cite{kafr03}.

Here we would like to point out some effects which have been overlooked
in the present literature. The main interest in the exponent $c$ is that
of modeling the DNA melting curves and the melting process typically
takes place at around $80^\circ - 90^\circ$ C (see Figs. \ref{eifg4},
\ref{LAMC1} and \ref{ecoli_fig}). Therefore, in order to discuss the
applicability of the higher loop exponent $c\approx 2.1$ to real DNA
sequence, it is the difference in persistence length of dsDNA and
ssDNA in this range of temperatures which should be investigated.
At room temperatures the persistence lengths are $\xi_{\rm ds}(T =
20^\circ {\rm C}) \approx 500$ {\AA}, while $\xi_{\rm ss}(T = 20^\circ
{\rm C}) \approx 40$ {\AA} corresponding to roughly $100$ bps and $8$
bps, respectively \cite{mark95}.  Both conformational fluctuations and
electrostatic interactions contribute to the persistence length of dsDNA
in solution \cite{odij77}. In the limit of high salt concentrations
electrostatic interactions are totally screened and the persistence
length is expected to scale as function of the temperature as in the
classical wormlike chain model \cite{doi87}, i.e. $\xi = \kappa/T$,
with $\kappa$ temperature independent. This formula implies a reduction
of about $20\%$ of the dsDNA persistence length in the melting region
compared to the room temperature value. However the assumption that the
electrostatic interactions are totally screened may not be fully justified
for the range of salt concentrations used here ($[Na]=0.1$M or lower).
Thus the $20\%$ should be rather considered as an upper bound.

A more relevant effect for the reduction of the dsDNA persistence length
is the proliferation of small denaturated bubbles within a dsDNA segment.
Within the model considered here the
characteristic lengths close to melting, as the loop and double helix
segment lengths, scale as $1/\sigma$. As pointed out before, a large
cooperativity (small $\sigma$), would imply typically long loops and long
double helical segments. However, the conclusion that short (i.e. $\sim
10$ bps) loops would be totaly suppressed due to the small $\sigma$,
is too simplistic. Several studies showed that the nearest neighbor
model largely underestimates the opening probability for small loops
(see the discussion in \cite{wart85} and references therein), so it cannot 
be a quantitatively reliable tool at too short length scales.
Such short loops are present in real DNA samples.

\begin{figure}[t]
\includegraphics[height=6.0cm]{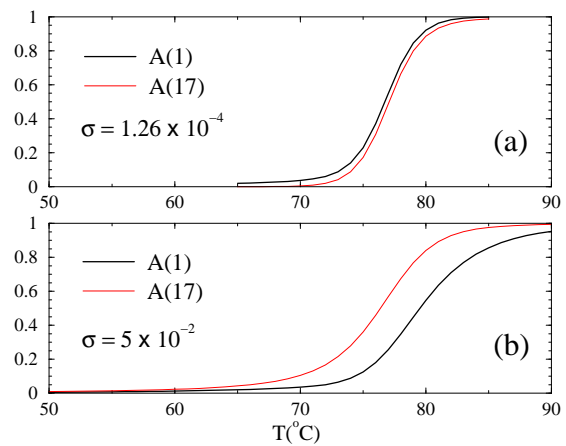}
\caption{Probability of finding the $i=1$ and $i=17$ in an open state
plotted as function of the temperature for the sequence M$_{\rm 18}$
of Ref. \cite{alta03} for $\sigma = 1.26 \times 10^{-4}$ (a) and
$\sigma = 1.26 \times 10^{-1}$ (b).
 }
\label{bonnet_fig}
\end{figure}

We can demonstrate this effect explicitly by analyzing the experimental
melting curves recently obtained \cite{alta03} in a study of dsDNA
bubble dynamics. This study was performed on short sequences (of about
$30$ bps) containing an internal AT region surrounded on both sides by
short CG clamps. Experiments show that the inner AT region melts at
temperatures in which the CG edges are still bound \cite{alta03}.

In Fig. \ref{bonnet_fig} we present the calculated melting curves
for the probabilities of having the base pair $i=1$ and $i=17$ in
an open state for the sequence $M_{18}$ of Ref. \cite{alta03},
which can be directly compared with the experimental results as
both quantities $A(1)$ and $A(17)$ have been measured
experimentally through fluorescence measurements \cite{alta03}
(the base pair $i=17$ is in the middle of the AT-region). As the
sequence is very short the calculated melting curves are not
sensitive to a change in $c$. For $\sigma = 1.26 \times 10^{-4}$
(Fig. \ref{bonnet_fig}(a)) the typical value for the cooperativity
parameter used previously in the paper, our calculations show that
$A(1)$ and $A(17)$ are at all temperatures very close to each
other, which implies that no loops are formed and that the
sequence rather melts from edge openings, although the CG
edges are energetically more stable than the inner AT region. This
is an effect of the small value of $\sigma$ used in the
calculation. In order to verify this we have plotted in Fig.
\ref{bonnet_fig}(b), just as an illustration,  the same quantities
with $\sigma = 5 \times 10^{-2}$.  The reduced cooperativity allows
for the formation of loops and now produces results closer to
what is observed in experiments (see Fig. 2 of Ref. \cite{alta03}).
Thus, despite that the small cooperativity parameter ($\sigma \sim
10^{-4}$) correctly describes long loop ($ > 100$ bps) formations
in DNA melting, small loops ($\sim 10$ bps) openings in AT-rich
domains are still possible. The importance of small bubbles
formation in DNA oligomers melting has been recently emphasized in
an analysis of the melting of DNA oligomers \cite{zocc03}.

As the ssDNA has, at $90^\circ$ C, an estimated persistence length
corresponding to roughly $5$ bps, we expect that the opening of a small
loop of about $15$ bps would be sufficient to decorrelate completely
the two dsDNA segments at its two sides. To provide an estimate of
the persistence length of the dsDNA one would need to know the average
density of small loops and their probabilities, which depends on the
sequence composition. It is however conceivable that this effect would
make the dsDNA at $80 - 90 ^\circ$C much more flexible, compared to its
room temperature behavior, so that the use of the loop embedded exponent
for $c$ is justified.

\section{Conclusion}
\label{sec:conclusion}

In this paper we have analyzed the effect of reparametrizing the
loop weight contribution, in the calculation of DNA melting curves for
sequences of a broad range of lengths, up to the full genome of the E-coli
($5.6 \times 10^6$ bps).  Using the closure exponent for a loop embedded
in a chain ($c = 2.15$) we found that in order to reproduce correctly
the existing experimental data and melting curves one needs to increase
the cooperativity parameter $\sigma$ by about one order of magnitude. An
increase of the cooperativity parameter, which is the weight associated
with the interruption of an helix to form a loop, implies that loops
are more probable within the chain.

It is clear that a simultaneous change of the loop exponent $c$ and of
$\sigma$ cannot reproduce two perfectly overlapping melting curves.
We found that rescaling $\sigma$ by about one order of magnitude
together with a change in $c$ from $1.75$ to $2.15$ produces typically
very small shifts in peak positions ($\sim 0.1^\circ$) and heights.
Accurate melting experiments would be able to distinguish between the
two choices and fix unequivocally both $c$ and $\sigma$. In any case our
analysis indicates that the best samples where to test the above effects
are sequences of intermediate lengths ($\approx 10^3-10^4$ bps). We
showed that in these sequences rather large loops may be formed and that
the associated melting peaks are extremely sensitive to a change in loop
parameters $c$ and $\sigma$.  The disadvantage of shorter sequences
is that they predominantly melt through end openings (unless they are
designed to do otherwise, as in the example of the preceding section).
The melting curves of very long sequences, as we showed for E. Coli,
are only weakly affected by a change in the parameters $c$ and $\sigma$.

{\bf Acknowledgment.} We thank I. Diesinger and E. Meese for their help
in finding suitable sequences for melting analysis.  Discussions with
A. Hanke, R. Metzler and D. Mukamel are gratefully acknowledged.

\end{document}